\documentclass[aps,pre,10pt,twocolumn,noshowpacs,longbibliography,superscriptaddress]{revtex4-1}
\usepackage{amsfonts}
\usepackage{amsmath}
\usepackage{amssymb}
\usepackage{graphicx}
\usepackage{hyperref}
\begin{document}
%
%
%
%
\title{On the accuracy of message-passing approaches to percolation in complex networks}
\author{Antoine \surname{Allard}}
\affiliation{D\'epartement  de  Physique,  de  G\'enie  Physique,  et  d'Optique, Universit\'e  Laval,  Qu\'ebec  (Qu\'ebec),  Canada  G1V 0A6}
\affiliation{Centre de mod\'elisation math\'ematique, Universit\'e  Laval,  Qu\'ebec  (Qu\'ebec),  Canada  G1V 0A6}
\author{Laurent \surname{H\'ebert-Dufresne}}
\affiliation{Department of Computer Science and Vermont Complex Systems Center, University of Vermont, Burlington VT}
\affiliation{D\'epartement  de  Physique,  de  G\'enie  Physique,  et  d'Optique, Universit\'e  Laval,  Qu\'ebec  (Qu\'ebec),  Canada  G1V 0A6}
\date{\today}
\begin{abstract}
  The Message-Passing Approach (MPA) is the state-of-the-art technique to obtain quasi-analytical predictions for percolation on real complex networks.  Besides being intuitive and straightforward, it has the advantage of being mathematically principled: it is exact on trees, while yielding generally good predictions on networks containing cycles as do most real complex networks.  Here we show that the MPA does not perform its calculations on some ill-defined tree-like approximation of the network, as its formulation leads to believe, but rather considers a random network ensemble in which the original network is cloned and shuffled an infinite number of times. We conclude that the fact that the MPA is exact on trees does not imply that it is nearly exact on tree-like networks. In fact we find that the closer a non-tree network is to a tree, the worse the MPA accuracy becomes.
\end{abstract}
\maketitle
%
%
%
%
%
%
Message passing---also refered to as the cavity method or belief propagation---is a general class of inference methods used to solve a wide range of problems, from optimization in unsupervised learning to mean-field approximations in statistical physics~\cite{Huang2016,Mezard2009,Zdeborova2016}.  An important feature of these methods is that they provide exact predictions when the underlying structure of the problem at hand can be represented as a tree (e.g., Ising chain, Bethe lattice) while still offering surprisingly good approximated solutions for structures containing loops~\cite{Weiss2000}.

In the context of network science, message-passing approaches have been used to shed a new light on several canonical problems such as epidemic spreading~\cite{Altarelli2014a,Karrer2010,Lokhov2014,Lokhov2015,Shrestha2015,Wilkinson2014}, resting neuronal activity~\cite{Peraza-Goicolea2019arXiv}, opinion dynamics~\cite{Lokhov2015,Shrestha2014,Wang2019b}, community detection~\cite{Zhang2014}, complex contagion~\cite{Gleeson2018}, spectral analysis~\cite{Newman2019,Newman2019arXiv}, spin models~\cite{DelFerraro2015,Lokhov2015} and percolation~\cite{Bianconi2016,Bianconi2017,Bianconi2018,Cellai2016,Karrer2014,Kuhn2017,Morone2015,Radicchi2016,Radicchi2015,Timar2017}.  While these different approaches rely on the assumption that the original network is locally tree-like, they are mostly intended for real complex networks that typically contain short loops.  A common assumption is therefore that the closer a network structure is to a tree, the more accurate the predictions from message passing should be.

We argue that this assumption, although seemingly reasonable at first sight, can lead to misguided conclusions.  Focusing on bond percolation, we show that message passing is in fact always exact, but that it performs its calculations on a well-defined random network ensemble whose structure may significantly differ from the original network.  Particularly, for networks whose structure is \textit{almost} a tree, message passing may predict behaviours that greatly diverge from the ones obtained by numerical simulations---the difference, in some cases, being as drastic as the emergence of a phase transition.
%
%
%
%
\begin{figure*}[t]
  \centering
  \includegraphics[width = 0.495\linewidth]{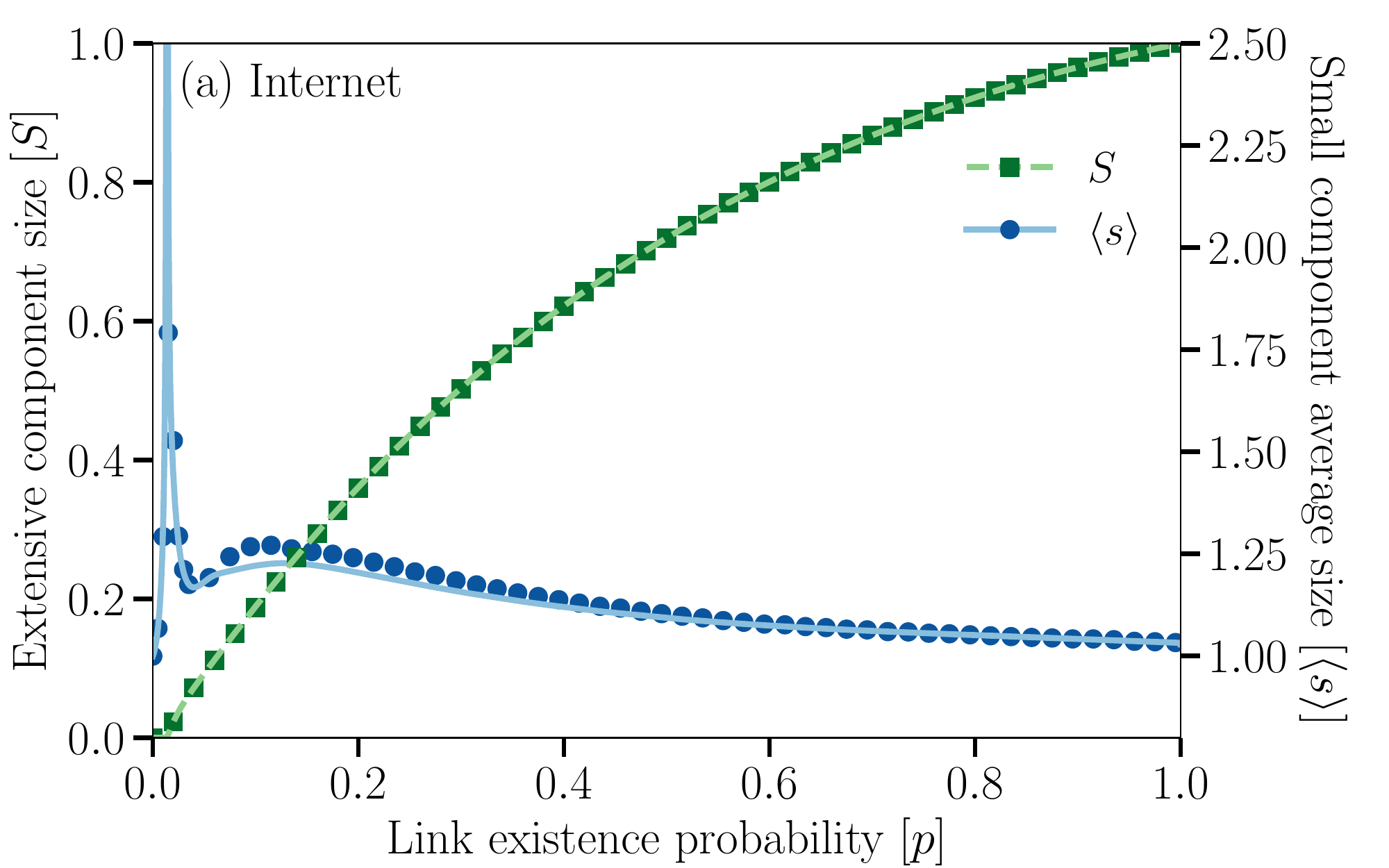}
  \includegraphics[width = 0.495\linewidth]{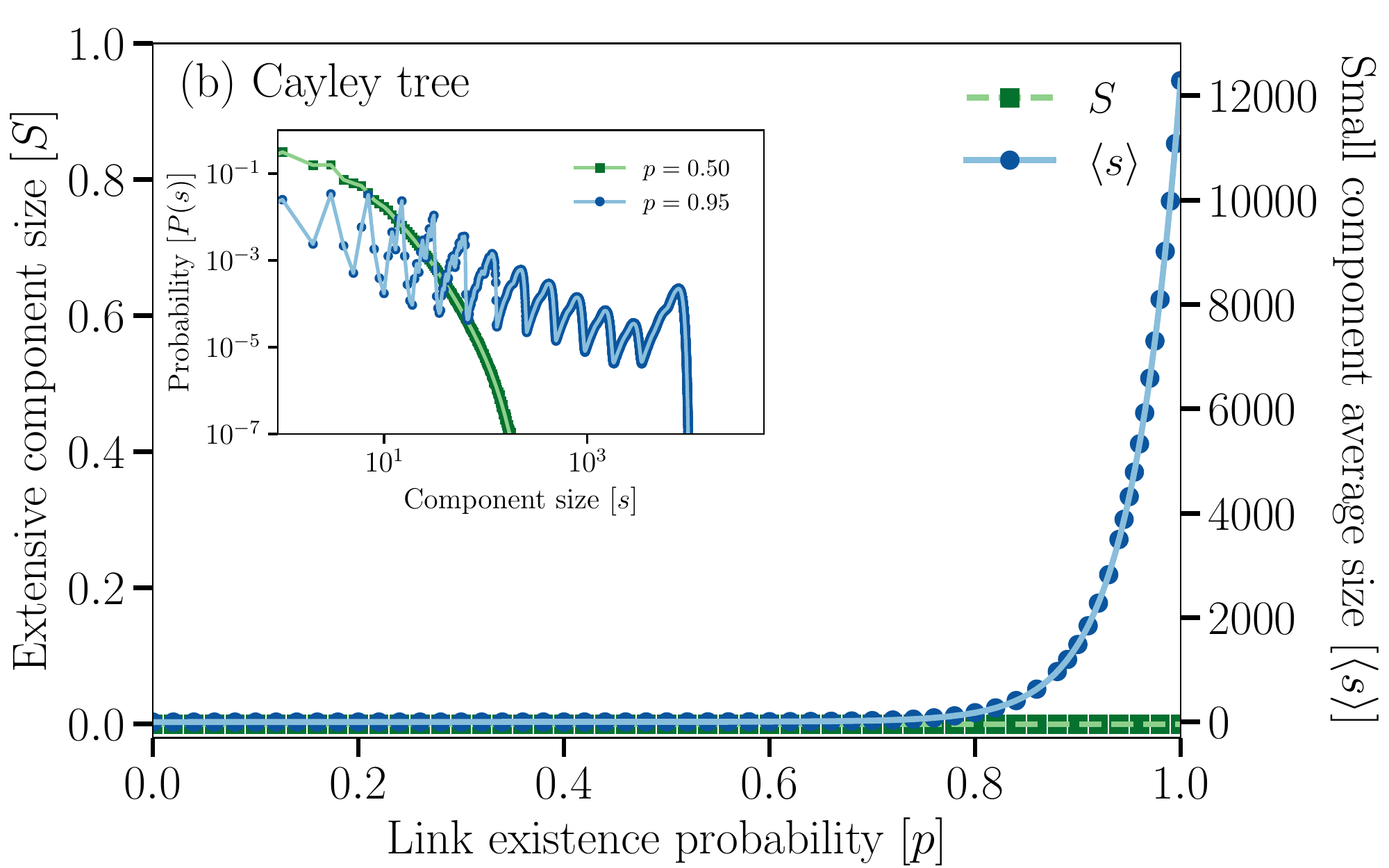}
  \caption{Comparison between the predictions of the MPA (lines) and the results obtained by numerical simuations (symbols) for the relative size of the extensive component ($S$) and the average size of the small component to which a randomly chosen node belongs ($\langle s \rangle$).  The theoretical predictions were obtained by solving Eq.~\eqref{eq:S} and Eq.~\eqref{eq:average_size_small_component}.  The numerical results were obtained by first removing each link in the original network with probability $1 - p$, and then by mesuring the size of the component a randomly chosen node belonged to.  The symbols show to the value of this size averaged over a large number of realizations (approx. $7.7\times10^7$ and $2.7\times10^6$ respectively).
  (a) Snapshot of the Internet~\cite{Leskovec2005} containing $N=11461$ nodes with an average local clustering coefficient $\langle c \rangle \simeq 0.35$.  A component was considered small if it contained a number of nodes smaller than $N/100$.
  (b) Cayley tree with a coordination number equal to 3 and 12 generations ($N=12286$).  The agreement between the predictions of the MPA (lines) and the results of numerical simulations (symbols) was possible only by considering every component as small, thereby confirming the absence of a phase transition.  Inset: Distribution $\{P(s)\}$ for $p=0.5$ and $p=0.95$ predicted by Eq.~\eqref{eq:small_component_size_distribution} (lines) compared with the results of numerical simulations (symbols, approx. $5\times10^9$ and $3\times10^9$ realizations respectively).  Note that Eq.~\eqref{eq:small_component_size_distribution} predicts a discrete distribution; lines are used to guide the eye.}
  \label{fig:prediction_mpa_phase_transition}
\end{figure*}

\textit{A message-passing approach to percolation}---%
%
Bond (site) percolation is a simple stochastic process related to the connectivity of networks of which links (nodes) have been independently removed with probability $1-p$.  It is a canonical problem of network science since it connects critical phenomena theory and statistical mechanics with many applied problems such as disease propagation or the robustness of real complex systems.

The message-passing approach (MPA)~\cite{Karrer2014} stands out from other analytical approaches in network science in that it uses an extensive description of the network as its input information---as opposed to an intensive description (e.g., degree distribution~\cite{Newman2001}, degree correlations~\cite{Vazquez2003}, onion decomposition~\cite{Allard2019}).  More precisely, it uses an adjacency list containing the set of the neighbours of each node $i$, which we will note $\mathcal{N}_i$.  The number of elements in the adjacency list is therefore equal to twice the number of links, $M$, which is an extensive quantity (proportional to the number of nodes, $N$, for sparse networks).

To predict the outcome of percolation, the MPA defines $u_{ij}$ as the probability that following the link from node $i$ to node $j$ does not lead to the extensive ``giant'' component.  This situation occurs if the link has been removed (probability $1-p$), or if the node at the other end of the link does not itself lead to the extensive component through its other neighbours. Altogether, this can be written as
\begin{equation} \label{eq:u_ij}
  u_{ij} = (1 - p) + p \prod_{l \in \mathcal{N}_j\backslash i} u_{jl} \ ,
\end{equation}
for $i, j = 1, \ldots, N$, and where $\mathcal{N}_j\backslash i$ corresponds to the neighbours of node $j$ excluding node $i$.  Notice that this previous expression relies on the assumption that the probabilities associated with the outgoing links (i.e., $u_{jl}\ \forall\ l \in \mathcal{N}_j\backslash i$) are independent for any node $j$, which is only true for a tree.  Notice also that $u_{ij} \neq u_{ji}$ in general so that the probability for a link to not lead to the extensive component depends on the direction in which it is followed.

Having solved Eq.~\eqref{eq:u_ij} for every ordered pairs $ij$, the probability that a node $i$ is not part of the extensive component is simply the probability that none of its neighbors lead to it, $\prod_{j \in \mathcal{N}_i} u_{ij}$, which again relies on the independence assumption of the $\{u_{ij}\}$ for any given node $i$.  The expected size of the extensive component, $S$, is therefore the average probability that any given node belongs to it
\begin{equation} \label{eq:S}
  S = \frac{1}{N} \sum_{i=1}^{N}  \Bigg( 1 - \prod_{j\in\mathcal{N}_i} u_{ij} \Bigg) \ .
\end{equation}

The value of $p$, $p_\mathrm{c}$, at which the extensive component emerges is
\begin{equation} \label{eq:p_c}
  p_\mathrm{c} = \left[ \lambda_\mathrm{max}(\mathbf{B}) \right]^{-1} \ ,
\end{equation}
where $\lambda_\mathrm{max}(\mathbf{B})$ is the leading eigenvalue of the Hashimoto (or nonbacktracking) matrix $\mathbf{B}$~\cite{Karrer2014}.

A slight modification to this formalism allows the calculation of the size distribution of the non-extensive ``small'' component to which a randomly chosen node belongs, $\{P(s)\}$.  Note that this quantity is related to the size distribution of the small components, noted $\pi_s$, via $P(s)\propto s \pi_s$ since a component of size $s$ contains $s$ times more nodes than a component of size 1, and is therefore $s$ times more likely to be the component to which a randomly chosen node belongs.

To compute $P(s)$, we substitute the probabilities $\{u_{ij}\}$ by the function $\{H_{ij}(z)\}$ generating the distribution of the number of nodes that will eventually be reached by following the link from node $i$ to node $j$.  Equation~\eqref{eq:u_ij} thus becomes
\begin{equation} \label{eq:H_ij}
  H_{ij}(z) = (1 - p) + p z \prod_{l \in \mathcal{N}_j\backslash i} H_{jl}(z) \ ,
\end{equation}
where the extra factor $z$ counts the contribution of node $j$ to the size of the component.  The distribution of the size of the small component to which a randomly chosen node belongs is then generated by averaging the number of nodes that can eventually be reached from any node $i$ in the network
\begin{equation} \label{eq:K}
  K(z) = \frac{z}{N} \sum_{i = 1}^{N} \prod_{j \in \mathcal{N}_i} H_{ij}(z) \ ,
\end{equation}
where the extra factor $z$ counts the initial node $i$.  The probability $P(s)$ is finally obtained by virtue of Cauchy's differentiation formula~\cite{Newman2001}
\begin{equation} \label{eq:small_component_size_distribution}
  P(s) = \left. \frac{1}{s!} \frac{\partial^{s} K(z)}{\partial z^s} \right|_{z=0} = \frac{1}{2 \pi i} \oint_\Gamma \frac{K(z)}{z^{s+1}} dz \ ,
\end{equation}
where $\Gamma$ is generally chosen to be the unit circle $|z|=1$.  Note that whenever $S > 0$, this last distribution must be divided by $1 - S$ to be properly normalized.

The first moment of the distribution $\{P(s)\}$, $\langle s \rangle$, can be obtained without solving Eqs.~\eqref{eq:H_ij}--\eqref{eq:small_component_size_distribution} explicitly.  Noting that the first moment of a distribution corresponds to the first derivative of its generating function, we get
\begin{equation} \label{eq:average_size_small_component}
  \langle s \rangle = \frac{K^\prime(1)}{K(1)}
                    = 1 + \frac{1}{N(1\!-\!S)} \sum_{i = 1}^{N} \sum_{j \in \mathcal{N}_i} H^\prime_{ij}(1) \prod_{l \in \mathcal{N}_i \backslash j} u_{il}
\end{equation}
where we noted ${X}^\prime(1) = \left. \frac{\partial X(z)}{\partial z} \right|_{z = 1}$.  The ``unnormalized'' expected size of the small component reached by following the link from node $i$ to node $j$, $H^\prime_{ij}(1)$, is the solution of
\begin{equation} \label{eq:H_ij_prime}
  H^\prime_{ij}(1) = p \prod_{l \in \mathcal{N}_j\backslash i} u_{jl}
                   + p \sum_{l \in \mathcal{N}_j\backslash i} H^\prime_{jl}(1) \prod_{n \in \mathcal{N}_j\backslash i,l} u_{jn} \ ,
\end{equation}
which is obtained by differentiating Eq.~\eqref{eq:H_ij}.  We refer the reader to Refs.~\cite{Karrer2014,Newman2001} for further details on the derivation of these equations.
%
%
%
%
\begin{figure}[b]
  \centering
  \includegraphics[width = \linewidth]{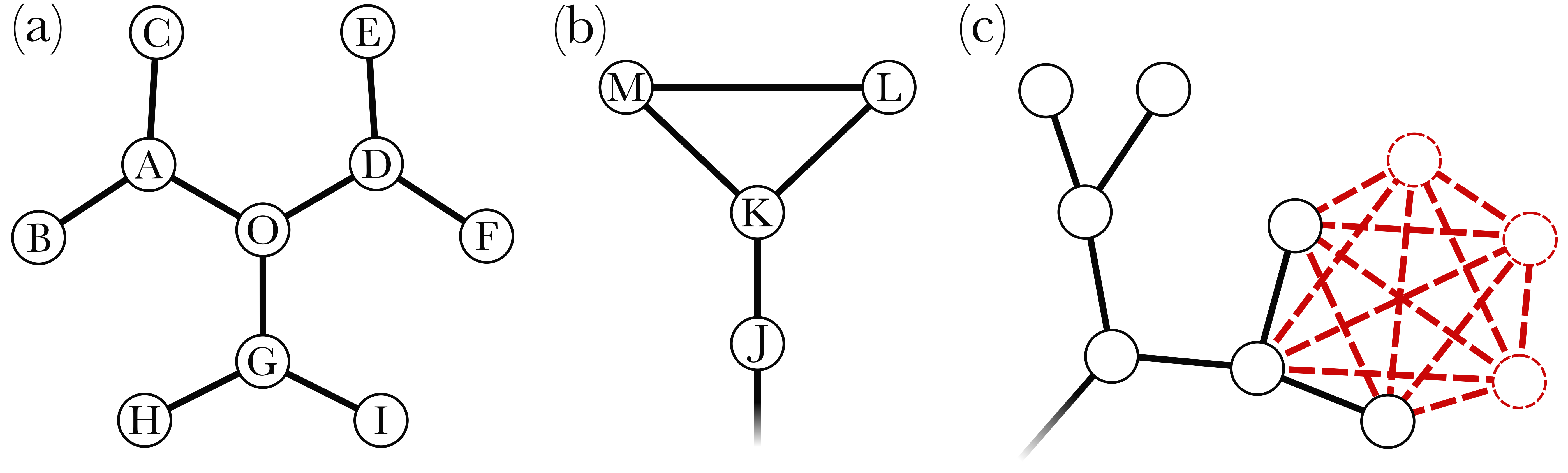}\vspace{\baselineskip}\\\vspace{\baselineskip}
  \includegraphics[width = \linewidth]{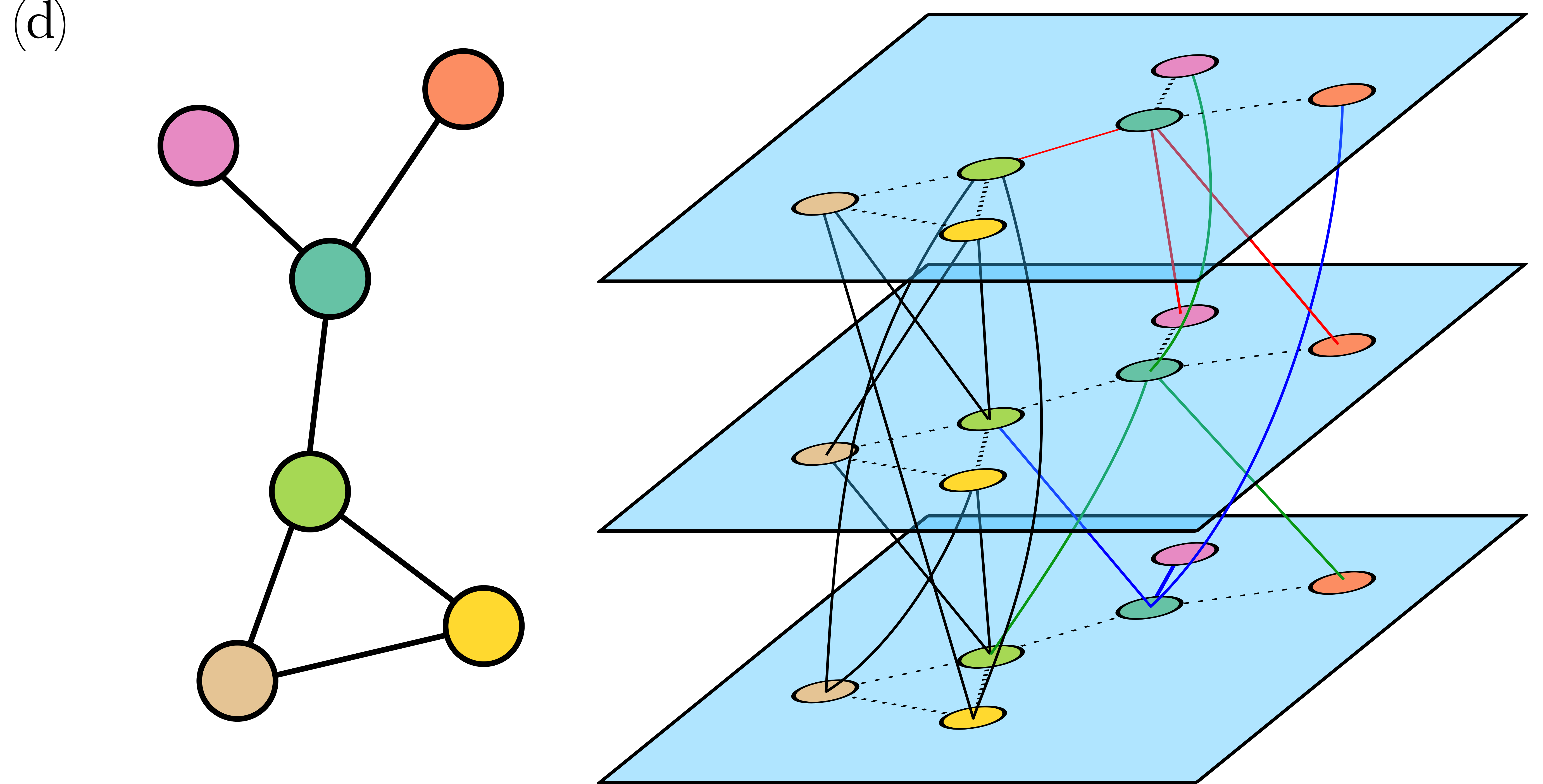}\vspace{\baselineskip}\\
  \caption{%
    (a)--(c) The simple networks considered in the main text to illustrate the innerworkings of the MPA.  The network shown on (c) is the same Cayley tree as on Fig,~\ref{fig:prediction_mpa_phase_transition}(b) to which 3 nodes and 13 links have been added to create a clique of 6 nodes at the end of a branch (shown using red dashed lines).
    (d) Illustration of the $L$-cloning procedure introduced in Ref.~\cite{Faqeeh2015} with $L=3$ for the small network shown on the left.  Further details are provided in the main text.%
  }
  \label{fig:cartoons}
\end{figure}

\textit{Percolation on a tree}---%
%
The fundamental assumption behind Eqs.~\eqref{eq:u_ij}--\eqref{eq:H_ij_prime} is that the ``state'' of two links stemming out of a same node are independent: the probabilities $\{u_{ij}\}$ are uncorrelated for any node $i$.  The same goes with the functions $\{H_{ij}(z)\}$).  Consequently, we expect the predictions of the MPA to be exact on a tree.  Nevertheless, as shown on Fig.~\ref{fig:prediction_mpa_phase_transition}(a), the MPA can be strikingly accurate even for networks whose local structure contains a high density of loops [Fig.~\ref{fig:prediction_mpa_phase_transition}(a) shows the results for a network whose average local clustering coefficient is close to 0.35].

However, a drastically different phenomenology is observed when the MPA is used on a Cayley tree of a similar size.  Figure~\ref{fig:prediction_mpa_phase_transition}(b) supports that predictions are exact, as expected, but the network does not percolate at $p=0.5$ as suggested by the mean-field analysis on a Bethe lattice with a same coordination number~\cite{Christensen2005}.  In fact, there is no phase transition at all (i.e., $S=0$ for all $p \in [0,1]$), and the only way to reach an agreement between the predictions of the MPA and the results of numerical simulations of bond percolation on the Cayley tree is by considering every components as ``non-extensive'' regardless of their size.  In other words, the MPA considers any component on a tree as a small component, in sharp contrast with Fig.~\ref{fig:prediction_mpa_phase_transition}(a) even though both networks contain roughly the same number of nodes.  The inset on Fig.~\ref{fig:prediction_mpa_phase_transition}(b) further corroborates that no phase transition occurs at $p=0.5$ through the absence of scale invariance in the distribution $\{P(s)\}$.  It also confirms that the increase of $\langle s \rangle$ before reaching $p=1$ on Fig.~\ref{fig:prediction_mpa_phase_transition}(b) does not indicate the imminent onset of a phase transition (recall that $\langle s \rangle$ is related to the second moment of the size distribution of small components, $\pi_s$, and is expected to diverge at the phase transition~\cite{Christensen2005}).  The behaviour of $\langle s \rangle$ as $p$ approaches 1 instead reflects the fact that an increasingly larger non-extensive component exists, and that this non-extensive component consists in the whole tree at $p = 1$.

Let us consider the tree of Fig.~\ref{fig:cartoons}(a) to understand the reasons behind this absence of phase transition.  From Eq.~\eqref{eq:u_ij}, it is clear that $u_{ij}=1$ whenever $j$ is a leaf of the tree (nodes $B$, $C$, $E$, $F$, $H$ and $I$) since these links lead to a dead end and therefore cannot lead to the extensive component when followed in the ``outward'' direction.  In turn, this observation implies that $u_{ij}=1$ also for the three links leaving $O$ toward $A$, $D$ and $G$ since they can only lead to links that will not lead to the extensive component.  The same reasoning can be applied once more for the links reaching $O$ from $A$, $D$ and $G$, and then again for the links leaving the leaf nodes toward $A$, $D$ and $G$.  We therefore conclude that $u_{ij}=1$ for every ordered pair $ij$, and consequently that the MPA will predict $S=0$ for every $p\in [0,1]$ (no extensive component).  In other words, the MPA ``sees'' the finiteness of the tree, and concludes adequately that no phase transition can occur.  This argument is straightforward to generalize to any network whose maximal $k$-core is the 1-core (i.e., any tree), and consequently that the MPA will predict the absence of a phase transition for any network without cycles.
%
%
%
%
\begin{figure}[t]
  \centering
  \includegraphics[width = \linewidth]{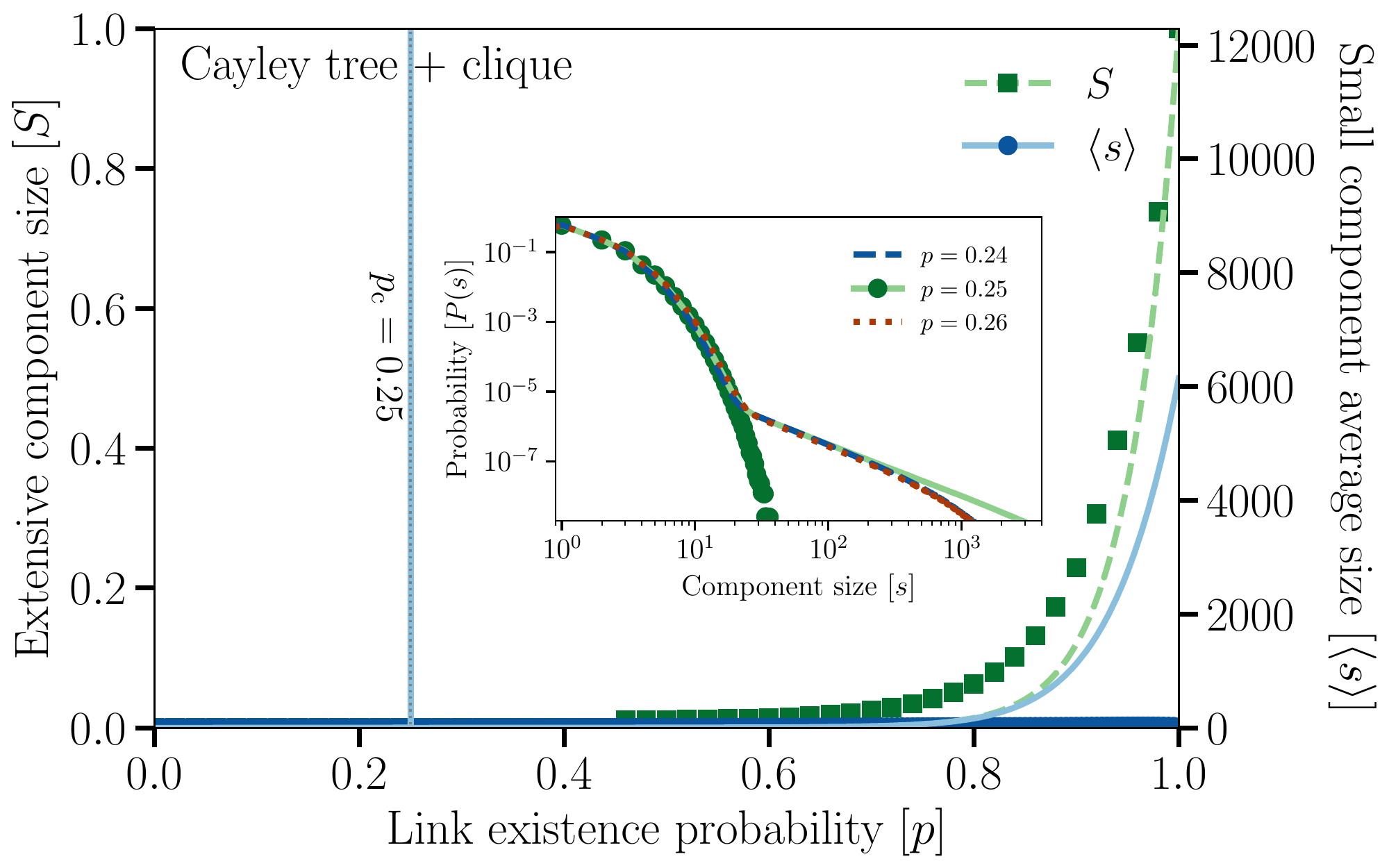}
  \caption{Comparison between the predictions of the MPA (lines) and the results obtained by numerical simuations (symbols) for the modified Cayley tree showed on Fig.~\ref{fig:cartoons}(c) ($N = 12289$).  Most of the details provided in the caption of Fig.~\ref{fig:prediction_mpa_phase_transition} apply here except for the following ones.  A component was considered small if it contained a number of nodes smaller than $N/100$.  The symbols were obtained by averaging over the results of approx. $2.7\times10^7$ realizations.  The value of the percolation threshold, $p_\mathrm{c} = 0.25$, is shown as a thin vertical dotted line and was obtained by solving Eq.~\eqref{eq:p_c}.
  Inset: Distribution $\{P(s)\}$ for $p=0.24$, $p=0.25$ and $p=0.26$ predicted by Eq.~\eqref{eq:small_component_size_distribution} (lines).  Only the results of numerical simulations for $p=0.25$ (symbols, approx. $1.5\times10^{10}$ realizations) are displayed to reveal the absence of scale-free behavior in the distribution, thereby showing that the phase transition is an artefact of the MPA only.  Note that Eq.~\eqref{eq:small_component_size_distribution} predicts a discrete distribution; lines are used to guide the eye.}
  \label{fig:finite-size}
\end{figure}

\textit{What is the MPA percolating on?}---%
%
If the MPA ``sees'' the finiteness of any tree, how come it does not appear to ``see'' the finiteness of the network used on Fig.~\ref{fig:prediction_mpa_phase_transition}(a)?  Indeed, contrarily to Fig.~\ref{fig:prediction_mpa_phase_transition}(b), we clearly observe the signature of a phase transition: the emergence of the extensive component (order parameter) and the divergence of $\langle s \rangle$. Altogether, this suggests that the MPA somehow considered the limit of infinite size.

The difference between these two scenarios is rooted in the presence of cycles, and Fig.~\ref{fig:cartoons}(b) allows us to illustrate how it is so.
Focusing on the link from node J to node K, one could reasonably expect the MPA to conclude that $u_{JK} = 1$ since the link leads to a dead end, albeit in the shape of a triangle, and therefore cannot lead to the giant component.  However, working out Eq.~\eqref{eq:u_ij} for this specific example yields a very different conclusion.  We first see that $u_{JK}$ depends on $u_{KL}$ and $u_{KM}$ as
\begin{equation}
  u_{JK} = (1 - p) + p u_{KL} u_{KM} \ ,
\end{equation}
and, by applying Eq.~\eqref{eq:u_ij} three times, we then find the following expressions for $u_{KL}$ and $u_{KM}$
\begin{subequations} \label{eq:mpa_on_triangle}
\begin{align}
  u_{KL} & = \!(1\!-\!p)\!+\!p\Big[ (1\!-\!p)\!+\!p\big[ (1\!-\!p)\!+\!pu_{KJ}u_{KL} \big] \Big]     \label{eq:mpa_on_triangle_u_KL} \\
  u_{KM} & = \!(1\!-\!p)\!+\!p\Big[ (1\!-\!p)\!+\!p\big[ (1\!-\!p)\!+\!pu_{KJ}u_{KM} \big] \Big]\ .  \label{eq:mpa_on_triangle_u_KM}
\end{align}
\end{subequations}
From these equations, we first observe that both $u_{KL}$ and $u_{KM}$ depend on $u_{KJ}$, thereby providing a simple illustration of how the presence of cycles contradicts the independence assumption on which Eqs.~\eqref{eq:u_ij}--\eqref{eq:H_ij_prime} are based.  Moreover, we see that $u_{JK}$ depends on $u_{KJ}$ and that the occupation of the link when crossed from node K to J is independent of its existence when crossed from J to K (i.e., the MPA redraws the existence of a link every time it is crossed, as if it were a different link).  Finally, we see that both values, $u_{KL}$ and $u_{KM}$, depend on themselves due to the periodicity of the cycle.  While there exist several ways to solve Eqs.~\eqref{eq:mpa_on_triangle}, the general approach to solve Eq.~\eqref{eq:u_ij} is by iterating them from random initial values until the $\{u_{ij}\}$ have converged to a satisfying precision~\cite{Karrer2014}.  Applying this method to solve Eqs.~\eqref{eq:mpa_on_triangle}, we see that doing so would therefore be equivalent to reinjecting Eqs.~\eqref{eq:mpa_on_triangle} into themselves a (formally) infinite number of times.

Altogether, these observations suggest that, rather than solving percolation on the loop itself, the MPA will go through an \textit{infinite number of ``copies''} of a same cycle, each time reaching different, yet identical copies of nodes J, K and L, and each time redrawing whether any link exists or not.  In other words, solving Eq.~\eqref{eq:u_ij} is equivalent to unfolding any cycle into an ``infinite'' sub-network that preserves correlations between and beyond nearest neighbors while being devoid of any loops of finite length.  By unfolding the entire network into a new copy every time the MPA has to go through a loop, it is essentially cloning the network infinitely many times and randomizing every link $\{ij\}$ across the infinite numbers of copies of nodes $i$ and $j$.  The network that the MPA ``sees'' is therefore the same as the one defined by the $L$-cloning process introduced in Ref.~\cite{Faqeeh2015} in the limit $L \rightarrow \infty$.
%
%
%
%
\begin{figure}[b]
  \centering
  \includegraphics[width = \linewidth]{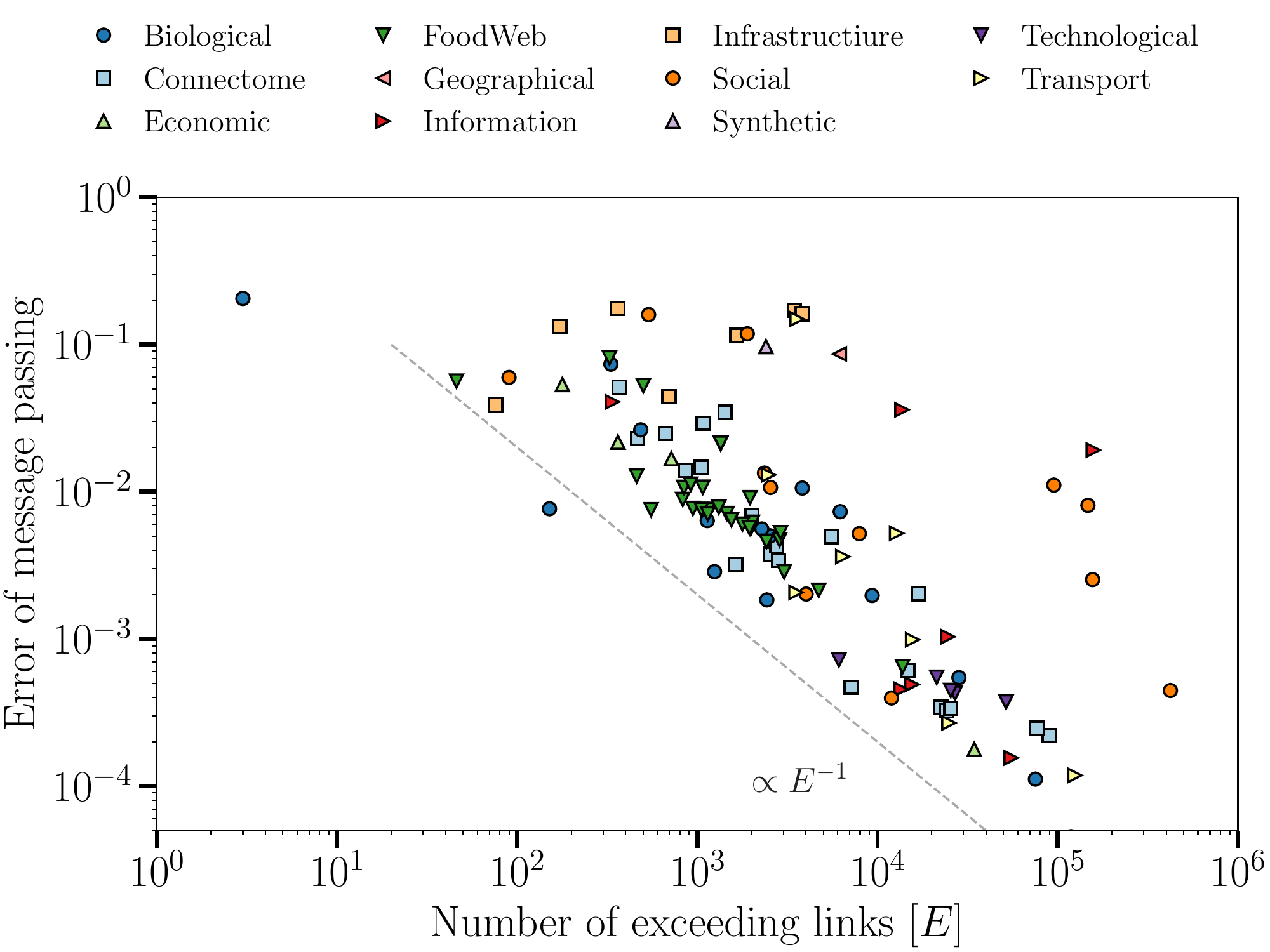}
  \caption{Scatter plot of the error made by the MPA on 111 network datasets (see Ref.~\cite{Allard2019} for a description of the datasets).  Error is measured by the absolute difference between the size of the extensive component predicted by the MPA [Eq.~\eqref{eq:S}] and its estimate obtained through numerical simulations.  The error is averaged over all values of the existence probability $p$ between 0 and 1 in increments of 0.005 (see Ref.~\cite{Allard2019} for further details).  Deviation from a perfect tree is measured by the number of exceeding links $E$ defined as the total number of links $M$ minus the number of links in a tree with the same number of nodes $N$ [i.e., $E=M-(N-1)$].  The error tends to be maximal for tree-like networks even though it is zero for trees.  A dashed line proportional to $E^{-1}$ provides a rough estimate of how the error of the MPA scales with $E$.}
  \label{fig:scatter}
\end{figure}

$L$-cloning was introduced to study the effect of clustering on various dynamical processes on networks and proceeds as follows.  It first makes $L$ identical copies of the original network, thereby creating $L$ copies of every nodes and every links (layers).  Then, for each connected pair of nodes in the original network, it shuffles the $L$ copies of the link across the different layers.  For instance, if nodes A and B are connected in the original network, then $L$-cloning creates the copies A\textsuperscript{(1)},\ldots, A\textsuperscript{($L$)}, B\textsuperscript{(1)},\ldots, B\textsuperscript{($L$)} and shuffles the links such that A\textsuperscript{(1)} may become connected to B\textsuperscript{($3$)}, A\textsuperscript{(6)} may become connected to B\textsuperscript{($2$)}, etc. As shown in Ref.~\cite{Faqeeh2015}, this proceedure defines an ensemble of random networks with the exact same degree distribution and degree-degree correlation between and beyond the nearest neighbors, but whose density of loops of any fixed length approaches zero for sufficiently large values of $L$. Figure~\ref{fig:cartoons}(d) provides a simple illustration of the $L$-cloning procedure.

The insight on the innerworkings of the MPA that $L$-cloning now provides allow us to explain the difference between the outcome of the two networks considered on Fig.~\ref{fig:prediction_mpa_phase_transition}.  Because a tree contains the minimum number of links to be connected (i.e., $M = N - 1$), shuffling links across layers will not merge the $L$ trees into one single large tree.  In other words, applying $L$-cloning to a tree will always yield $L$ identical copies of the original tree of \textit{finite} size $N$.  The MPA will therefore ``see'' the finiteness of the trees and will not predict any phase transition.

However, if a network contains at least a 2-core, then the $E \equiv M - (N - 1)$ exceeding links will connect the layers, and $L$-cloning will most likely generate one single connected network of size $LN$ if $E$ is moderately large (note that components in the original network, if any, will be preserved by $L$-cloning). Because it considers networks produced by $L$-cloning in the limit $L \rightarrow \infty$, the MPA will therefore ``see'' an infinite network on which a genuine phase transition is possible.

\textit{Infinite finite-size effects}---%
%
Let us now consider a slightly modified version of the Cayley tree used for Fig.~\ref{fig:prediction_mpa_phase_transition}(b) to test this hypothesis.  It consists of the same tree with the exception that 3 nodes and 13 links were added at the very end of one branch to complete a 6-node clique [see Fig.~\ref{fig:cartoons}(c)].

From the perspective of the numerical simulations, this addition should be insignificant---it is a mere perturbation in the periphery of the tree.  On the contrary, the $L$-cloning interpretation indicates that the MPA will instead ``see'' a network whose core consists in an infinite 5-regular random network (the unfolded clique) with 1/6 of its nodes having an extra link leading to a finite tree composed of 12283 nodes (there is an infinite number of finite trees).  The addition of the peripheral clique therefore induces a strong core-periphery structure with a phase transition at $p_\mathrm{c} = 1/4$ driven by the core.  This prediction is confirmed on Fig.~\ref{fig:finite-size} by the solution of Eq.~\eqref{eq:p_c}, by the position of the divergence of $\langle s \rangle$ as well as by the scale-free regime in the distribution $P(s)$.  Most importantly, when compared to the results of numerical simulations, we see that the addition of the small clique drastically impairs the accuracy of the MPA.  While not being a formal proof, our results nevertheless strongly support that the MPA considers the ensemble of random networks defined by $L$-cloning in the limit $L \rightarrow \infty$ instead of some ill-defined tree-like approximation of the original network.

This interpretation suggests that the MPA may therefore perform terribly as soon a few exceeding links are added on a tree since these extra links create loops and therefore connect the different cloned layers.  To further explore this conclusion, Fig.~\ref{fig:scatter} investigates the accuracy of the predictions of the MPA for the size of the extensive component on 111 network datasets with different levels of ``tree-likeness'' measured by the number of exceeding links $E \equiv M - (N - 1)$ ($E = 0$ for a perfect tree).  In line with our previous conclusions, we find that the MPA tend to perform worst on networks that deviate from trees by only a few links, and tend to perform much better on denser networks with more loops.  As a guide, we find that the error of the MPA scales roughly as $E^{-1}$.

\textit{Discussion}---%
%
Unveiling the random network ensemble on which the MPA performs its calculations sheds a clearer light on how the MPA is exact on trees, and most importantly how it suddenly ceases to be exact with the addition of loops.  More precisely, the abruptness of this transition means that its exactitude on trees does not imply that the MPA will be accurate on any tree-like networks, which is unfortunately a common assumption.  Instead, we find what could be described as a ``tree-like catastrophe'' where the MPA is most unreliable on trees with a handful of extra links, despite being exact on the original trees. However, as more links are added, the MPA surprisingly gains in accuracy as networks get denser and less tree-like; most likely from the feedback effect being diluted across many loops.

Our results also provide guidelines to anticipate and interpret the validity of the predictions of the MPA for networks containing cycles.  For instance, this perspective explains how the MPA captures percolation transitions smeared by meso-scale structures \cite{Hebert-Dufresne2018arXiv}.  Indeed, recent work showed that MPA can accurately estimate the percolation threshold of these smeared transitions although it cannot capture the mechanisms by which they appear.  For example, two weakly coupled modules with asymmetric sizes and densities can produce a scenario where the extensive component emerges in one module before transferring to the other.  Under the lense of message passing this scenario is impossible.  The modular structure is instead mapped to an effective core-periphery structure through the $L$-cloning procedure, and the percolation process always nucleates from the densest core before growing outward.  In fact, all networks with loops are mapped to an effective structure of nested cores through $L$-cloning.

Finally, we would like to stress that this work does not in any way discredit the usefulness of the MPA, but rather sheds an insightful light on its innerworkings to allow a more accurate interpretation of its predictions.  Most importantly, we warn against assuming that message passing and other related approaches should be accurate simply because a network is sparse.  While our conclusions only apply to percolation, they could be relevant to many of the other contexts in which message-passing techniques are confidently used to obtain accurate predictions.  We therefore hope this work will incite future similar enquiries, which will hopefully lead to a better understanding of the range of applicability of this powerful and versatile technique.

AA acknowledges financial support from the project Sentinelle Nord of the Canada First Research Excellence Fund and from the Natural Sciences and Engineering Research Council of Canada. LHD acknowledges support from the National Science Foundations Grant No. DMS-1829826 and the National Institutes of Health 1P20 GM125498-01 Centers of Biomedical Research Excellence Award.  This research was enabled in part by support provided by Westgrid and Calcul Canada.
%
%
%
%
%
%
%
%
%
%
%
%
\end{document}